\documentstyle[12pt,epsfig]{article}
\begin{document}
\begin{titlepage}
\begin{flushright}
ULB--TH 99/19\\
October 1999
\end{flushright}
\vspace*{1.6cm}

\begin{center}
{\Large\bf Neutrino gravitational lensing}\\
\vspace*{0.8cm}

R.~Escribano\footnote{Chercheur I.~I.~S.~N.}, 
J.-M.~Fr\`ere\footnote{Directeur de recherches du F.~N.~R.~S.},
D.~Monderen and V.~Van Elewyck\\
\vspace*{0.2cm}

{\footnotesize\it Service de Physique Th\'eorique, Universit\'e Libre de
Bruxelles, CP 228, B-1050 Bruxelles, Belgium}\
\end{center}
\vspace*{1.0cm}

\begin{abstract}
We study the lensing of neutrinos by astrophysical objects. At the difference of 
photons, neutrinos can cross a stellar core; as a result the lens quality 
improves. While Uranians alone would benefit from this effect in the Sun, 
similar effects could be considered for binary systems.
\end{abstract}
\end{titlepage}

\section{Introduction}
In this note we want to investigate the possibility of neutrino lensing by
astrophysical objects (stars, galaxies or rather galactic halo). At the difference 
of photons, neutrinos can cross even a star's core. As will be seen below this results in
a much better focalisation. On the other hand, the price to pay is the extreme difficulty
to detect neutrinos, and the comparatively poor angular resolution of "neutrino telescopes".
For this reason, the only thing we can hope for is a signal intensification, rather
than the spectacular photon lensing patterns.
The main equations are established in section 2, with details in the Appendix.
Section 3 deals with neutrino interactions in a stellar medium, and specify which
energy range can be studied in this way.
In section 4 we consider a number of situations and establish the
corresponding signal enhancement expected. 
Finally in section 5 we consider some practical examples.
It appears clearly that the Earth - Sun
distance is too small for a sizable effect to take place, (an observatory on Uranus 
would notice the enhancement of distant neutrino sources whenever
they are aligned with the Sun). We then consider the case of galaxies,
through their halo. We finally turn to binary systems.
We will not in this paper comment on the possible origin of 
energetic neutrinos,
but we remark that an interesting situation is met when one
of the companions focuses the neutrino flux originating from the other.

\section{Gravitational deflection of neutrinos}
\label{rafel}

In this section we study the deflection of neutrinos from straight-line motion
as they pass through a gravitational field produced by a compact object of 
mass $M$. 
We will distinguish two cases: when the neutrino flux passes far away
from the object (OUTside solution), a situation equivalent to the gravitational
lensing of photons, and, when the neutrino flux passes through the object
(INside solution). 
In the latter case, we consider three specific cases depending on the compact
object density profile: constant density (an academic but constructive
example), Gaussian distribution density (suitable for stars\footnote{
The density profile of stars is not exactly Gaussian but we use it here in order to
obtain simple analytical results. Such a description should be considered as a 
good approximation to the real case.}), and Lorentzian
distribution density (which could be associated to a galactic halo\footnote{
The Lorentzian profile behaves as $1/r^2$ for large $r$, in agreement with velocity
dispersion curves for galaxies and clusters.}).

We begin by calculating the trajectory of a massless neutrino (or with mass 
very small compared to its energy) in the Schwartzschild  metric under the 
assumption that $M/r$ is everywhere small along the trajectory \cite{Schutz}.
The equation of the orbit is\footnote{Geometrized units $c=G=1$ are used 
throughout the paper.}
\begin{equation}
\label{eqdiforbitOUTr}
\frac{d\phi}{dr}=
\frac{1}{r^2\sqrt{\frac{1}{b^2}-\frac{1}{r^2}\left(1-\frac{2M}{r}\right)}}\ ,
\end{equation}
where $b$ is defined as the impact parameter. 
Using the definition $u\equiv\frac{1}{r}$:
\begin{equation}
\label{eqdiforbitOUTu}
\frac{d\phi}{du}=
\frac{1}{\sqrt{\frac{1}{b^2}-u^2+2 M u^3}}\ .
\end{equation}
If we neglect the $u^3$ term in Eq.~(\ref{eqdiforbitOUTu}), all effects of $M$
disappear, and the solution is 
\begin{equation}
\label{eqorbitlineu}
r\sin(\phi-\phi_0)=b\ ,
\end{equation}
a straight line.
In the limit $M u\ll 1$, or $R_{\rm Sch.}\equiv 2M\ll b$, if we define 
$y\equiv u(1-M u)$, Eq.~(\ref{eqdiforbitOUTu}) becomes
\begin{equation}
\label{eqdiforbitOUTy}
\frac{d\phi}{dy}=
\frac{1+2 M y}{\sqrt{\frac{1}{b^2}-y^2}}+O(M^2 u^2)\ .
\end{equation}
Integrating Eq.~(\ref{eqdiforbitOUTy}) gives
\begin{equation}
\label{phiOUTy}
\phi_{\rm OUT}(y)=
\phi_0+\frac{2M}{b}+\arcsin(b y)-2M\sqrt{\frac{1}{b^2}-y^2}\ ,
\end{equation}
where the integration constant is defined as $\phi=\phi_0$ (with $\phi_0$
the incoming direction) when the initial trajectory has $r\rightarrow\infty$ 
(or $y\rightarrow 0$).
The particle reaches its smallest $r$ when $\frac{dr}{d\lambda}=0$:
\begin{equation}
\label{eqdifrminOUT}
\frac{dr}{d\lambda}=
E\sqrt{1-\left(1-\frac{2M}{r}\right)\frac{b^2}{r^2}}=0\ 
\Longrightarrow\ y_{\rm max}=\frac{1}{b}+O(M^2 u^2)\ .
\end{equation}
This occurs at the angle
\begin{equation}
\label{phiOUTymax}
\phi_{\rm OUT}(y=\frac{1}{b})=\phi_0+\frac{2M}{b}+\frac{\pi}{2}\ .
\end{equation}
It has thus passed through an angle 
$\frac{2M}{b}+\frac{\pi}{2}$ as it travels to its point of closest approach. 
By symmetry, it passes through a further angle of the same size as it moves 
outwards from its point of closest approach (see Ref.~\cite{Schutz}).
Then, the particle passed through a total angle of $\frac{4M}{b}+\pi$. 
If it were keeping  to a straight trajectory, this angle would be $\pi$, 
so the net deflection is
\begin{equation}
\label{phiOUTnet}
\Delta\phi_{\rm OUT}=\frac{4M}{b}\ .
\end{equation}

For the case of a neutrino flux passing through the object, we must first, in 
order to study the neutrino trajectory, look for the form of the space-time 
in the region inside the object. 
Here, we restrict ourselves to the case of static spherically symmetric 
space-times\footnote{Spherically symmetric space-times are reasonably simple,
yet physically very important, since very many objects of importance in 
astrophysics appear to be nearly spherical. A static space-time is defined to
be one in which we can find a time coordinate $t$ with two properties:
(i) all metric components are independent of $t$, and (ii) the geometry is 
unchanged by time reversal, $t\rightarrow -t$.}.
In that case, the most general metric is (see Ref.~\cite{Schutz} for details):
\begin{equation}
\label{metricIN}
ds^2=-e^{2\Phi(r)}dt^2+e^{2\Lambda(r)}dr^2+r^2 d\Omega^2\ ,
\end{equation}
where it is convenient to replace $\Lambda(r)$ by
\begin{equation}
\label{defLambda}
m(r)\equiv\frac{1}{2}r\left(1-e^{-2\Lambda}\right)\ 
\Longrightarrow\ g_{rr}=e^{2\Lambda}=\frac{1}{1-\frac{2m(r)}{r}}\ .
\end{equation}
For a static perfect fluid\footnote{A perfect fluid in relativity is defined
as a fluid that has no viscosity and no heat conduction in the momentarily
comoving reference frame (MCRF). 
A static fluid is a fluid that has no motion.}, 
Einstein equations imply
\begin{equation}
\label{Einsteineqs}
\begin{array}{l}
\frac{dm(r)}{dr}=4\pi r^2 \rho(r)\ ,\\[2ex]
\frac{d\Phi(r)}{dr}=\frac{m(r)+4\pi r^3 p(r)}{r(r-2m(r))}=
-\frac{1}{\rho(r)+p(r)}\frac{dp(r)}{dr}\ ,
\end{array}
\end{equation}
where $\rho(r), m(r)$ and $p(r)$ are, respectively, the density, mass and 
pressure of the object at a radius $r$.
For completeness, in the region outside the object we have $p=\rho=0$, then
\begin{equation}
\begin{array}{l}
m(r)=M={\rm const.}\ \Longrightarrow\ e^{2\Phi(r)}=1-\frac{2M}{r}\\[2ex]
\Longrightarrow\ 
ds^2=-\left(1-\frac{2M}{r}\right)dt^2+\frac{dr^2}{1-\frac{2M}{r}}+r^2 
d\Omega^2\ .
\end{array}
\end{equation}
In the region inside the object, exact solutions to the relativistic equations 
are very hard to solve analytically for a given equation of 
state \cite{Schutz}.
One interesting exact solution is the Schwarzschild constant-density interior 
solution, which we use here as an example of the framework needed to study 
other density profiles.

Inside $R$, where $R$ is the physical radius of the object, 
$\rho\neq 0, p\neq 0$.
For a constant density profile $\rho(r)=\rho$, the equation of the orbit is
(see the appendix for a detailed calculation)
\begin{equation}
\label{eqdiforbitINconstrhomaintext}
\frac{d\phi}{dr}=
\frac{\frac{3}{2}\sqrt{1-\frac{2M}{R}}-
      \frac{1}{2}\sqrt{1-\frac{2M}{R}\frac{r^2}{R^2}}}
     {\sqrt{1-\frac{2M}{r}\left(\frac{r}{R}\right)^3}}
\frac{1}{r^2\sqrt{\frac{1}{b^2}-\frac{1}{r^2}
\left(\frac{3}{2}\sqrt{1-\frac{2M}{R}}-
      \frac{1}{2}\sqrt{1-\frac{2M}{R}\frac{r^2}{R^2}}\right)^2}}\ ,
\end{equation}
and the net deflection is
\begin{equation}
\label{phiINnetconstrhomaintext}
\Delta\phi=
\left\{
\begin{array}{ll}
\frac{4M}{b} & {\rm if}\ b\geq R\\[2ex]
\frac{4M}{b}\left(1-\sqrt{1-\frac{b^2}{R^2}}\right)
+2\arcsin\left[\frac{b}{R}\left(1-\frac{M}{R}\right)\right]\\[2ex]
\ \ +\frac{3M}{R}\frac{b}{R}\sqrt{1-\frac{b^2}{R^2}}
    -2\arcsin\left\{\frac{b}{R}\left[1-\frac{3M}{2R}
     \left(1-\frac{b^2}{3R^2}\right)\right]\right\} & {\rm if}\ b<R
\end{array}
\right.
\end{equation}
where the outside solution is also included for completeness.

Next, we analyze the solutions for the Gaussian and Lorentzian distribution
densities. The Gaussian profile is a convenient approximation to the mass
distribution in stars, while the Lorentzian profile is valid for galactic 
halos.
In both cases, it is possible to neglect the pressure with respect to the mass
density, $p\ll\rho$ (see Ref.~\cite{Schutz} for the so-called Newtonian stars),
so we also have $4\pi r^3 p\ll m$.
Moreover, the metric must be nearly flat, so in Eq.~(\ref{defLambda}) we 
require $m(r)\ll r$. These inequalities simplify Eq.~(\ref{Einsteineqs}) to
\begin{equation}
\label{PhirlimNewton}
\frac{d\Phi(r)}{dr}=\frac{m(r)}{r^2}\ ,
\end{equation}
and the metric in Eq.~(\ref{metricIN}) to
\begin{equation}
\label{metriclimNewton}
\begin{array}{l}
e^{2\Lambda(r)}\simeq 1+\frac{2m(r)}{r}\hspace*{1cm} \mbox{and}\hspace*{1cm} 
e^{2\Phi(r)}\simeq 1+2\Phi(r)\\[2ex]
\Longrightarrow\ 
ds^2=-(1+2\Phi(r))dt^2+\left(1+\frac{2m(r)}{r}\right)dr^2+r^2 d\Omega^2\ .
\end{array}
\end{equation}

\begin{figure}[ht]
\centerline{\epsffile{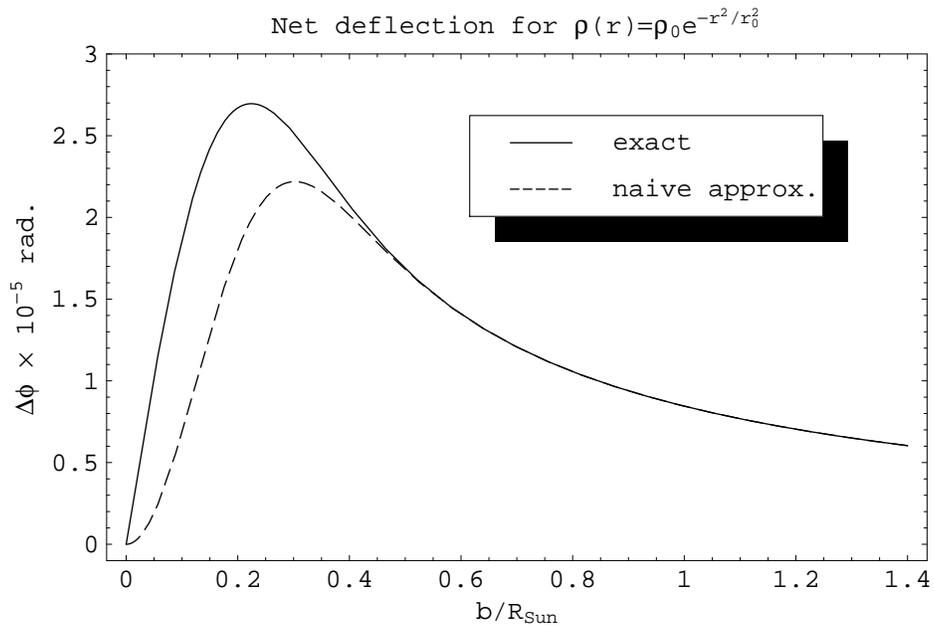}}
\caption{Net deflection $\Delta\phi$ as a function of the impact parameter $b$
for the case of a Gaussian density profile $\rho(r)=\rho_0 e^{-r^2/r_0^2}$ 
(see Eq.~(\protect\ref{phiINnetgaussianmaintext})). 
The na\"{\i}ve approximation of Eq.~(\protect\ref{naiveapproxgaussian}) is 
also included for comparison. In both curves, $M\equiv M_{\odot}=1.48\ {\rm km}$,
$R\equiv R_{\odot}=6.96\times 10^5\ {\rm km}$ and $r_0=0.2 R_{\odot}$ are used.} 
\label{plotgaussian} 
\end{figure}

For the case of a Gaussian density profile $\rho(r)=\rho_0 e^{-r^2/r_0^2}$, 
the equation of the orbit is (see appendix for details)
\begin{equation}
\begin{array}{rl}
\label{eqdiforbitINgaussianrhomaintext}
\frac{d\phi}{dr}= &
\left[1-\frac{M}{R}
\frac{e^{(R^2-r^2)/r_0^2}-1}
     {r_0/R\,e^{R^2/r_0^2}\sqrt\pi/2\,
      {\rm erf}\left(R/r_0\right)-1}\right]\nonumber\\[2ex]
\times & \frac{1}{r^2\sqrt{\frac{1}{b^2}-\frac{1}{r^2}
\left[1-\frac{2M}{R}
\frac{r_0/r\,e^{R^2/r_0^2}\sqrt\pi/2\,
      {\rm erf}\left(r/r_0\right)-1}
     {r_0/R\,e^{R^2/r_0^2}\sqrt\pi/2\,
      {\rm erf}\left(R/r_0\right)-1}\right]}}\ ,
\end{array}
\end{equation}
where the error function is defined as 
${\rm erf}(z)=\frac{2}{\sqrt\pi}\int_0^z dt\,e^{-t^2}$.
The net deflection is then
\begin{equation}
\label{phiINnetgaussianmaintext}
\Delta\phi=
\left\{
\begin{array}{ll}
\frac{4M}{b} & {\rm if}\ b\geq R\\[2ex]
\frac{4M}{b}\left(1-\sqrt{1-\frac{b^2}{R^2}}\right)
+\frac{4M}{b}\frac{r_0/R\,e^{R^2/r_0^2}\sqrt\pi/2}
{r_0/R\,e^{R^2/r_0^2}\sqrt\pi/2\,{\rm erf}\left(R/r_0\right)-1}\\[2ex]
\times\left[\sqrt{1-\frac{b^2}{R^2}}\,{\rm erf}\left(R/r_0\right)-
e^{-b^2/r_0^2}{\rm erf}\left(\sqrt{1-\frac{b^2}{R^2}}\frac{R}{r_0}\right)
\right] & {\rm if}\ b<R
\end{array}
\right.
\end{equation}
It is very interesting (because we are close to a real case) to compare the 
previous result with a na\"{\i}ve approximation where for a given impact 
parameter $b$ one studies the net deflection by a sphere of radius $b$ and 
identical density profile
\begin{equation}
\label{naiveapproxgaussian}
\begin{array}{rl}
\Delta\phi\mid_{\rm approx}\equiv & \frac{4m(b)}{b}\\[2ex]
= & \frac{4M}{R}e^{(R^2-r^2)/r_0^2}
\frac{r_0/r\,e^{r^2/r_0^2}\sqrt\pi/2\,
      {\rm erf}\left(r/r_0\right)-1}
     {r_0/R\,e^{R^2/r_0^2}\sqrt\pi/2\,
      {\rm erf}\left(R/r_0\right)-1}\ \ \ \ {\rm for}\ \ \ \ b<R
\end{array}
\end{equation}
In Fig.~\ref{plotgaussian}, we plot for comparison the exact result in 
Eq.~(\ref{phiINnetgaussianmaintext}) together with the na\"{\i}ve
approximation in Eq.~(\ref{naiveapproxgaussian}). 
We have taken the Sun as example of a typical star. Then, the mass and the
radius in Eqs.~(\ref{phiINnetgaussianmaintext},\ref{naiveapproxgaussian}) 
are fixed to $M\equiv M_{\odot}=1.48\ {\rm km}$ and 
$R\equiv R_{\odot}=6.96\times 10^5\ {\rm km}$. The parameter $r_0$ is taken to be
$r_0=0.2 R_{\odot}$. As it is seen from
Fig.~\ref{plotgaussian}, the maximal net deflection occurs at 
$b\simeq r_0$ and is 
$\Delta\phi(b\simeq r_0)\simeq 3.2\Delta\phi(b=R_{\odot})$,
showing that the lensing effect inside the star is bigger than the outside 
effect (except for $b\leq 0.04 R_{\odot}$). It is also shown that for the
na\"{\i}ve approximation the maximal deflection is displaced up to 
$b\simeq 1.5 r_0$ and its value is around 20\% smaller than for the exact 
result. So then, our calculation seems to be essential for a detailed 
analysis of the gravitational lensing of a neutrino beam by a star.

\begin{figure}[ht]
\centerline{\epsffile{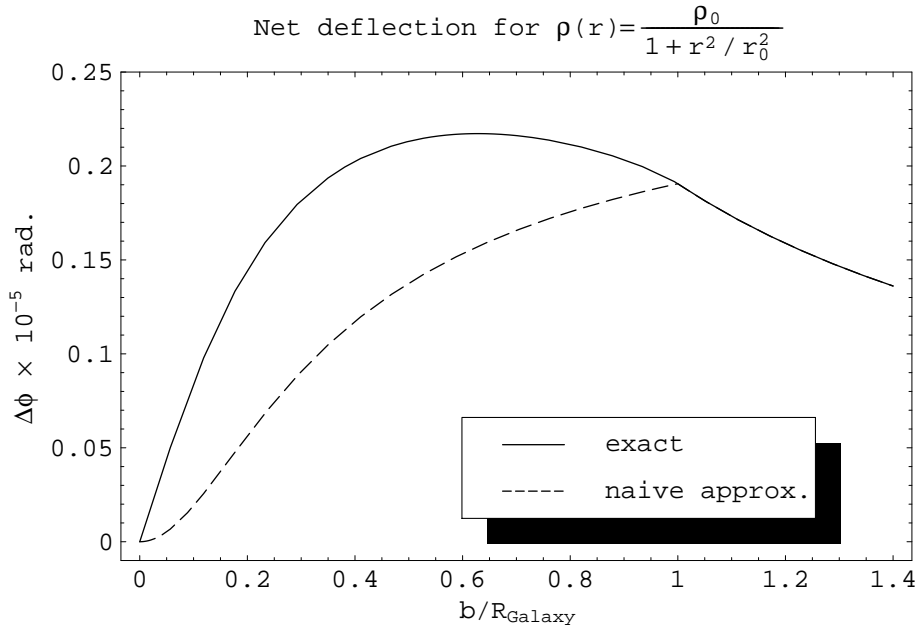}}
\caption{The same as in Fig.~\protect\ref{plotgaussian} but for a Lorentzian 
density profile $\rho(r)=\frac{\rho_0}{1+r^2/r_0^2}$.
$M\equiv M_{\rm Galaxy}=10^{12} M_{\odot}$, 
$R\equiv R_{\rm Galaxy}=100\ {\rm kpc}$ and $r_0=0.2 R_{\rm Galaxy}$ are used.} 
\label{plotlorentzian} 
\end{figure}

For the case of a Lorentzian density profile 
$\rho(r)=\frac{\rho_0}{1+r^2/r_0^2}$, the equation of the orbit is
\begin{equation}
\begin{array}{rl}
\label{eqdiforbitINlorentzianmaintext}
\frac{d\phi}{dr}= & 
\left[1+\frac{M}{R}\frac{\frac{1}{2}\log\left(\frac{r^2+r_0^2}{R^2+r_0^2}\right)}
                        {1-r_0/R\,\arctan(R/r_0)}\right]\nonumber\\[2ex]
\times & \frac{1}{r^2\sqrt{\frac{1}{b^2}-\frac{1}{r^2}
\left[1-\frac{2M}{R}
\frac{1-r_0/r\,\arctan(r/r_0)-\frac{1}{2} 
      \log\left(\frac{r^2+r_0^2}{R^2+r_0^2}\right)}
     {1-r_0/R\,\arctan(R/r_0)}\right]}}\ ,
\end{array}
\end{equation}
and the final deflection is
\begin{equation}
\label{phiINnetlorentzianmaintext}
\Delta\phi=
\left\{
\begin{array}{ll}
\frac{4M}{b} & {\rm if}\ b\geq R\\[2ex]
\frac{4M}{b}\left(1-\sqrt{1-\frac{b^2}{R^2}}\right)
-\frac{4M}{b}\frac{1}{1-r_0/R\,\arctan(R/r_0)}\frac{r_0}{R}\\[2ex]
\times\left[
\begin{array}{l}
\sqrt{1-\frac{b^2}{R^2}}\arctan(R/r_0)\\[2ex]
-\sqrt{1+\frac{b^2}{r0^2}}
\arctan\left(\frac{\sqrt{1-\frac{b^2}{R^2}}}{\sqrt{1+\frac{b^2}{r0^2}}}
             \frac{R}{r_0}\right)
\end{array}
\right] & {\rm if}\ b<R
\end{array}
\right.
\end{equation}
Here, it is also interesting to compare the previous result with the na\"{\i}ve
approximation for a Lorentzian density profile
\begin{equation}
\label{naiveapproxlorentzian}
\Delta\phi\mid_{\rm approx}=
\frac{4M}{R}\frac{1-r_0/r\,\arctan(r/r_0)}{1-r_0/R\,\arctan(R/r_0)}
\ \ \ \ {\rm for}\ \ \ \ b<R
\end{equation}
In Fig.~\ref{plotlorentzian}, we plot the exact result in 
Eq.~(\ref{phiINnetlorentzianmaintext}) and the na\"{\i}ve approximation in 
Eq.~(\ref{naiveapproxlorentzian}).
Here, we have taken $M\equiv M_{\rm Galaxy}=10^{12} M_{\odot}$ and
$R\equiv R_{\rm Galaxy}=100\ {\rm kpc}$ as an example for galaxies. 
In this case, with $r_0 = 0.2 R_{\rm Galaxy}$, the maximal net deflection occurs at
$b\simeq 3 r_0$ and is around 15\% bigger than the value
at $b=R_{\rm Galaxy}$, while for the na\"{\i}ve approximation the lensing 
effect inside the galactic halo is always smaller than the effect at
$b=R_{\rm Galaxy}$. Again, we think that a detailed calculation is convenient.

\begin{figure}[ht]
\centerline{\epsffile{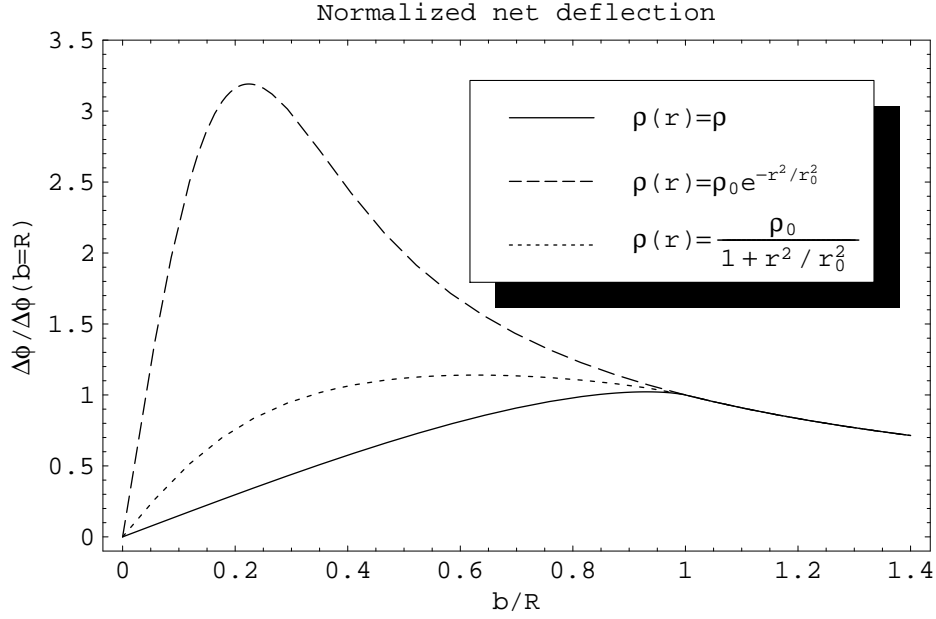}}
\caption{Normalized net deflection $\Delta\phi/\Delta\phi(b=R)$ as a function 
of the normalized impact parameter $b/R$ for a constant density profile
(solid line), a Gaussian density profile (dashed line) and a Lorentzian
density profile (dotted line). In the last two cases, $r_0$ is taken to be
$r_0=0.2 R$.} 
\label{plotalldeflectnor} 
\end{figure}

As a summary of this section, we plot in
Fig.~\ref{plotalldeflectnor} the normalized net deflection 
$\Delta\phi/\Delta\phi(b=R)$ as a function of the normalized impact parameter
$b/R$ for the three specific density profiles considered along the analysis:
constant density, Gaussian and Lorentzian distribution densities.
Such a normalization allows for a clear comparison among the three profiles
and is independent of the mass and physical radius of the compact object.
Only for the case of stars (Gaussian profile) the lensing effect at the
interior of the star is substantially amplified with respect to the outside
effect (the inside effect should be compared with the effect at $b=R$).
For galactic halos, the maximal inside net deflection is slightly bigger than
at $b=R$, while for an object of constant density the inside lensing effect
is always smaller than at $b=R$.

\begin{figure}[ht]
\centerline{\epsffile{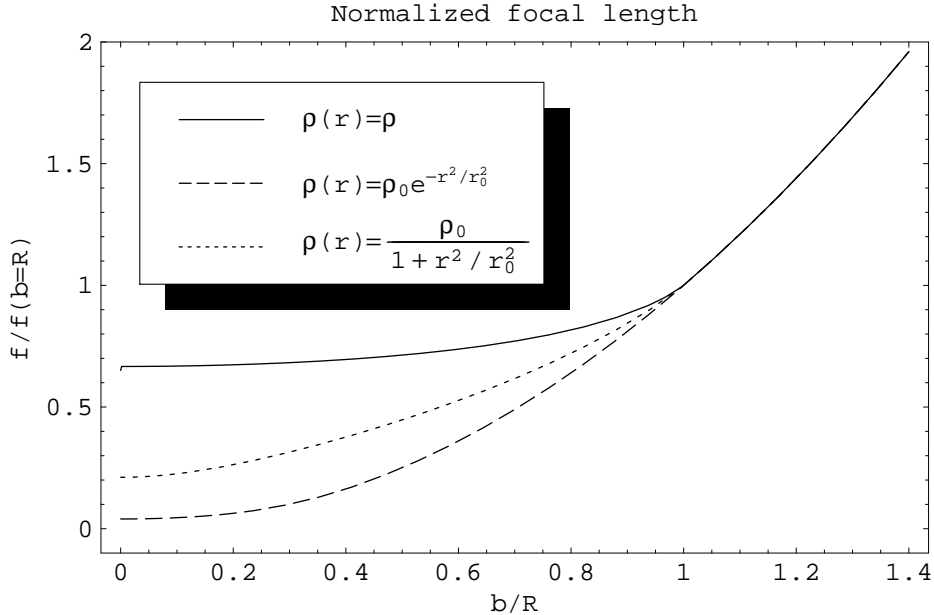}}
\caption{Normalized focal length $f/f(b=R)$ as a function 
of the normalized impact parameter $b/R$.} 
\label{plotallfocallengthnor} 
\end{figure}

Finally, we plot the focal length $f$ as a function of the
impact parameter $b$. The focal length is defined as the distance at which the
lens focuses the signal (see Sec.~\ref{tools} for details)
\begin{equation}
\label{focallength}
\Delta\phi=\frac{b}{f(b)}\ .
\end{equation}
In Fig.~\ref{plotallfocallengthnor}, the (normalized) focal lengths for the
three different profiles are drawn.
At $b\ll R$
\begin{equation}
\label{focallengthb0}
\frac{f(b\ll R)}{f(b=R)}=\left\{
\begin{array}{ll}
\frac{2}{3} & {\rm for}\ \ \rho(r)=\rho\\[2ex]
\frac{r_0/R\,e^{R^2/r_0^2}\sqrt\pi/2\,{\rm erf}\left(R/r_0\right)-1}
     {e^{R^2/r_0^2}\sqrt\pi/2\,{\rm erf}\left(R/r_0\right)}\frac{r_0}{R}
& {\rm for}\ \ \rho(r)=\rho_0 e^{-r^2/r_0^2}\\[2ex]
2\frac{1-r_0/R\,\arctan(R/r_0)}{\arctan(R/r_0)}\frac{r_0}{R}
& {\rm for}\ \ \rho(r)=\frac{\rho_0}{1+r^2/r_0^2}
\end{array}
\right.
\end{equation}
We postpone to Sec.~\ref{tools} (where focal lengths are discussed in
detail) the comments about the quality of the different gravitational lenses
considered here.
 
\section{Neutrino absorption}
While neutrinos are able to travel {\it across} the massive object, we have to consider their 
interactions with the matter inside; these will indeed 
reduce the neutrino flux and thus the efficiency of the signal amplification 
gained through lensing.  
At the energies of interest for cosmic 
neutrinos ($ E \geq 1-10 \,  GeV $), we have deep inelastic scattering off 
individual nucleons. Charged currents will cause the conversion of neutrinos
into charged leptons, which will later be absorbed or decay, while
neutral interactions will deflect them by angles in general large compared to the lensing effect.
In both cases, the interacting neutrinos will
be lost for lensing amplification.

 Calling \( \sigma \) the interaction cross-section 
(which depends on the neutrino energy) and \( N(x) \) the number density of scatterers along the
 neutrino path, we have the well-known formula for an infinitesimal flux variation due to interaction
with matter:
\begin{equation}
\label{pi}
d\Phi = N(x) .\sigma (E_{\nu}) .\Phi . dx
\end{equation}
The neutrino flux attenuation after passing through a layer of matter of depth $x$ is thus given by:
\begin{equation}
\label{pro}
\Phi (x,E_{\nu}) = \Phi_0 e^{-\sigma (E_{\nu}) \int_0^x N(x')dx'}
\end{equation}
Here we deal with objects having a spherical symmetry, so it is easier to express 
Eq.~(\ref{pro}) in terms of the radial variable \( r = \sqrt{x^2 + b^2} \), \(b\) being
 the impact parameter  of the neutrino respect to the center of the object. We then get
\begin{equation}
\label{proba}
\Phi (b,E_{\nu}) = \Phi_0 e^{-2\sigma (E_{\nu}) \int_b^R N(r)\frac{r dr}{\sqrt{r^2 - b^2}}}
\end{equation}
where $R$ is the radius of the object.
To solve this equation, we need to collect information on the relevant cross-sections, 
the possible mass density profiles and the composition of the object (that is, how to relate mass 
density to number density).\\

\subsection{Neutrino cross-sections}

At current accelerator energies, 
neutrino-proton and neutrino-neutron scattering differ (as the valence quark content is different), as
do neutrino and antineutrino interactions. From experimental results \cite{zp1,zp2},
 we adopt the following approximations :
\begin{equation}
\label{sigmalow}
\begin{array}{ll}
\sigma^{\nu p}_{tot}= 0.89\ \sigma^{\nu N}_{CC} \ \ & \ \  \sigma^{\bar{\nu} p}_{tot}=
 1.70\  \sigma^{\bar{\nu} N}_{CC} \\[2ex]
\sigma^{\nu n}_{tot}= 1.70 \ \sigma^{\nu N}_{CC}\ \  & \ \  \sigma^{\bar{\nu} n}_{tot}=
 1.03 \ \sigma^{\bar{\nu} N}_{CC} 
\end{array}
\end{equation}
using the tables of values for neutrino-isoscalar nucleon from Quigg, Reno {\it et al.}. \cite{GQRS} 

Above $10^6 $ GeV all these interactions become identical as sea quarks largely dominate. We may thus
directly use the values for scattering on an isoscalar nucleon (again from \cite{GQRS}) :
\begin{equation}
\label{sigmahigh}
\sigma^{\stackrel{\left( - \right) }{\nu} p}_{tot} \ = \ \sigma^{\stackrel{(-)}{\nu} n}_{tot}
\ =\ \sigma^{\stackrel{(-)}{\nu} N}_{tot}
\end{equation} \\
Strictly speaking we should interpolate between those two situations using appropriate
structure functions, however, in view of the results, the above approximations are sufficient.

Due to the smallness of the ratio $\frac{m_e}{m_{nucleon}}$, neutrino-electron interactions
will be neglected compared to neutrino-nucleon ones, except
in the particular case of the Glashow resonance
$\bar{\nu}_e e^- \rightarrow W^- \rightarrow anything$, that occurs at an energy
$E_{\bar{\nu}_e } = 6.3 $ PeV. At that energy we have \cite{GQRS95}
\begin{equation}
\sigma^{\bar{\nu}_e e^-}  \simeq 5 \times 10^{-31} \ cm^2
\end{equation}

Having taken into account the various components of the cross section, we
rewrite the formula~(\ref{proba}) giving the transmission
 probability of the neutrino as:
\begin{equation}
\label{probaX}
P_T(b,E_{\nu}) = exp\ [-2 \sum_X \sigma^{\nu X} (E_{\nu}) \ 
\int_b^R N_X(r) \ \frac{r dr}{\sqrt{r^2 - b^2}} \  ]
\end{equation}
where X stands for each kind of scatterer. The number 
densities $N_X(r)$ depend on the composition and mass density profile of the lens in the following
 obvious way :
\begin{equation}
\label{N}
N_X(r) =  \frac{{\cal M}_X}{M_X} \ \rho(r)
\end {equation}
where ${\cal M}_X$ and $M_X$ are respectively the fraction in mass and the molar mass of constituent
 X; $\rho$ is the mass density of the lens.
Let us now investigate the probability of transmission for the two particular cases of physical
relevance that have been discussed in the previous section, i.e. stars and galaxies :

\subsection{Neutrino transmission through a star}

All the following calculations apply in a good approximation to all the Sun-like stars 
\footnote{Neutrino flux attenuation in the Sun has also been considered in \cite{IngThun}.}.
We will suppose that the star is made up of 25 \% of Helium and 75 \% of Hydrogen in
 number; we then get the following relations between number and mass densities :
\begin{eqnarray}
N_{He}(r) &=& 0.142 \, mol \,\, (\frac{\rho(r)}{1 gr})       \\
N_{H}(r) &=& 0.427 \, mol \,\, (\frac{\rho(r)}{1 gr})              
\end{eqnarray}
that we used in our numerical calculations. We adopted for the mass density profile a Gaussian 
distribution (normalized to the total mass $M$, in the approximation $r_0 < 0.4R$) that reads:
\begin{equation}
\label{R}
\rho(r) = \frac{1}{\pi^{3/2} erf(\frac{R}{r_0})}\, \frac{M}{r_0^3}\,\, e ^{-(\frac{r}{r_0})^2}
\end{equation}
in function of the total mass M and radius R of the star, for a  distribution in which most of
 the object's mass is concentrated in a sphere of radius $r_0$ (we have normalized to the total
 mass).
We performed detailed calculations in the case $r_0 = \frac{R}{5}$; using Eqs.~(\ref{N}) and (\ref{R}) 
the transmission probability 
$P_T = \frac{\Phi}{\Phi_0}$ (\ref{probaX})
then becomes
\begin{eqnarray}
P_T(b,E_{\nu}) &=& exp\,\, \{-\frac{250}{\pi^{3/2}erf(5)} \, \frac{M}{R^3}\,\,
[ \sigma^{\nu He}\,\frac{{\cal M}_{He}}{M_{He}}+\sigma^{\nu H}\,\frac{{\cal M}_H}{M_H}] \times \nonumber %%@
\\[1ex]  
               && \,\,\,\,\,\, \times \,\,\int_b^R e ^{-(\frac{5r}{R})^2}
\frac{r dr}{\sqrt{r^2-b^2}}\ \} \nonumber \\[2ex]
               &=& exp\,\, \{-\frac{25}{\pi}\, \frac{M}{R^2}\,\, \, 
\frac{erf(5\sqrt{1-\frac{b^2}{R^2}})}{erf(5)}  \,\,e^{-(\frac{5b}{R})^2} \,\times \nonumber \\[1ex]
               && \,\,\,\,\,\, \times \,\,[ \sigma^{\nu p}\,(2 \frac{{\cal M}_{He}}{M_{He}}
+\frac{{\cal M}_H}{M_H})\, + \sigma^{\nu n} \, (2 \frac{{\cal M}_{He}}{M_{He}} )\,]\,\, \}
\end{eqnarray}

\begin{figure}[ht]
\centerline{\epsffile{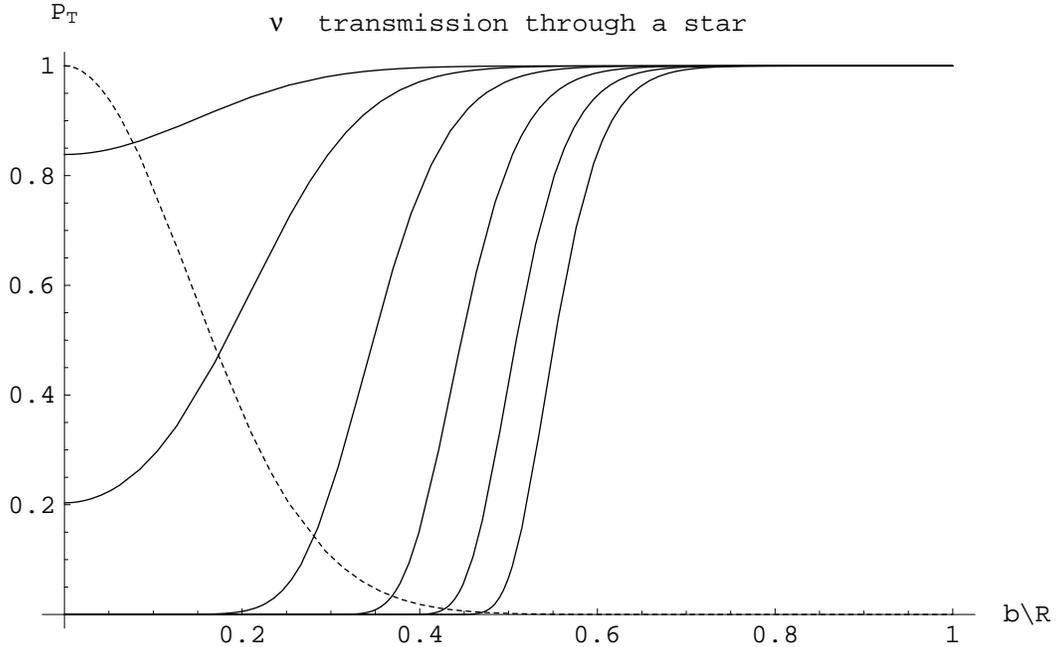}}
\caption{{Transmission probability in function of b, the impact parameter, for a neutrino energy
 $E_{\nu} =$ 10 GeV, $10^2$ GeV, $10^3$ GeV, $10^4$ GeV, $10^5$ GeV, $10^6$ GeV
 (solid lines, from left to
 right); the Gaussian
density profile $\rho(r) = e ^{-(\frac{5 r}{R})^2}$ is the dotted line.} }
\label{nu}
\end{figure}

\begin{figure}[ht]
\centerline{\epsffile{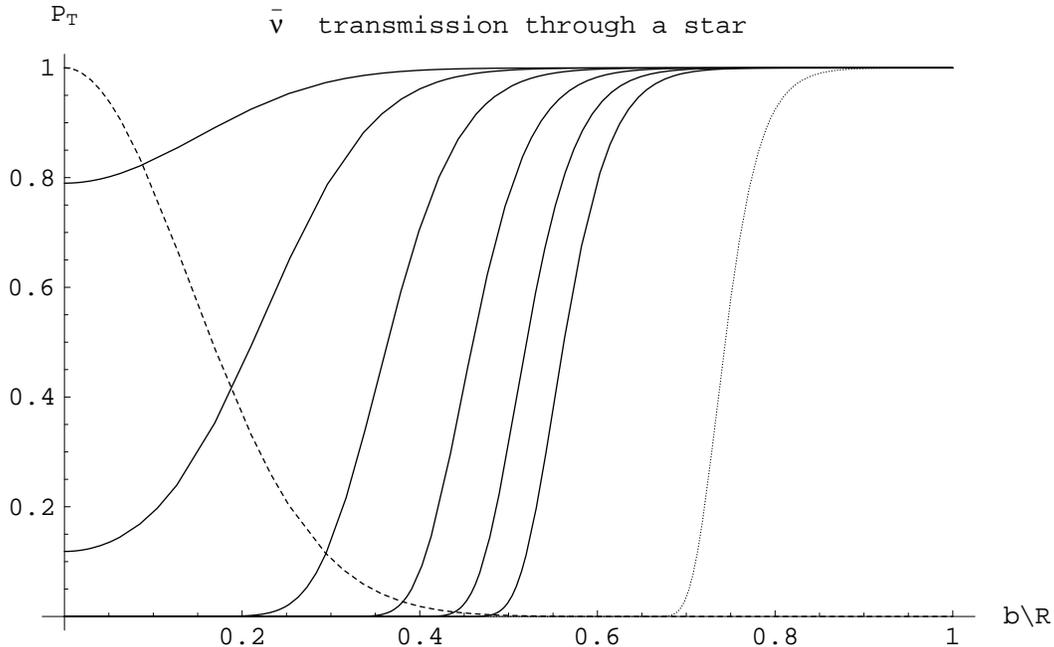}}
\caption{ {Transmission probability in function of b, the impact parameter,
for an antineutrino energy
 $E_{\bar \nu} =$ 10 GeV, $10^2$ GeV, $10^3$ GeV, $10^4$ GeV, $10^5$ GeV, $10^6$ GeV
 (solid lines, from left to right). Dashed
line on the right is for $E_{\bar{\nu}} = 6.3 $ PeV and corresponds to $\bar{\nu}_e e^-$ scattering.}}
\label{antinu}
\end{figure}

We present in Fig.~\ref{nu} and~\ref{antinu} the plots of the transmission probability
of neutrinos across the Sun as a function of $b/R_\odot$; we have added 
the density profile on the same figure.  For antineutrinos we have drawn
 the curve at $E_{\bar{\nu}_e}
 = $ 6.3 PeV where  $\bar{\nu}_e e^-$ is dominant; in this particular case, transmission probability can 
be written as
\begin{eqnarray}
P_T(b,6.3 PeV) &=& exp\,\, \{-\frac{25}{\pi}\, \frac{M}{R^2}\,\, \, 
\frac{erf(5\sqrt{1-\frac{b^2}{R^2}})}{erf(5)}  \,\,e^{-(\frac{5b}{R})^2} \,\times \nonumber \\
               && \,\sigma^{\bar{\nu} e^- }(6.3 PeV)\,[\, 2 \frac{{\cal M}_{He}}{M_{He}}
+\frac{{\cal M}_H}{M_H}\,]\ \}
\end{eqnarray} 
as electron and proton densities are supposed to be equal in stars.

As we see from Fig.~\ref{nu} and~\ref{antinu}, neutrinos with energies of 1 TeV or more are completely %%@
absorbed or
scattered away in the core of the star. Even at 100 GeV the interactions are frequent enough
 to significantly attenuate the flux of neutrinos coming out of the star.
Unfortunately, the zone where lensing could be the most efficient (as the star's core acts as a
"real lens") is thus ruled out because of flux attenuation for neutrinos of energy 100 GeV or more.
But we can see that under 100 GeV, interactions of neutrinos inside the star have nearly no 
incidence on the outcoming flux, which is totally recovered after lensing by the star.

\subsection{Neutrino transmission through a galaxy}
Neutrinos passing through a galaxy may interact either with its visible matter or with the
 surrounding halo of massive relic neutrinos. These last interactions become significant only
for incoming neutrinos at ultra high energies  (of the order of $10^{19} eV$)\cite{roulet} and we may %%@
neglect them in our present
frame of work. 

Concerning the visible part of the Galaxy, a rough estimate gives an average density of stars of
1 $pc^{-3}$, which corresponds to a negligible probability
for the neutrino to encounter a star during its passage through the galaxy, even in the worst case 
if it traverses the whole disk and the bulge. 

We thus conclude that the passage of neutrinos through the galaxy won't decrease their flux, and hence
 do not affect the lensing effect.

\section{Signal enhancement}
\label{tools}

\subsection{Geometry}

\subsubsection{General relations}
While a significant enhancement will only be reached for well - aligned lens,
detector and source, we need to consider unaligned elements when dealing
with extended sources or when taking into account the detector size.
We follow closely  \cite{gl}, and remind the basic geometric relations

\begin{eqnarray}
\label{angles}
\beta = \theta - \frac{D_{LS}}{D_{SO}} \alpha
\end{eqnarray}

where $\alpha$ \footnote{So far, it has been called $\Delta \phi$.} is the deflection angle; $\beta$, 
the angular position of the source and $\theta$, the angular position
of the image. See Fig.~(\ref{def}).

\begin{figure}[ht]
\label{def}
\epsfxsize=10cm
\centerline{\epsffile{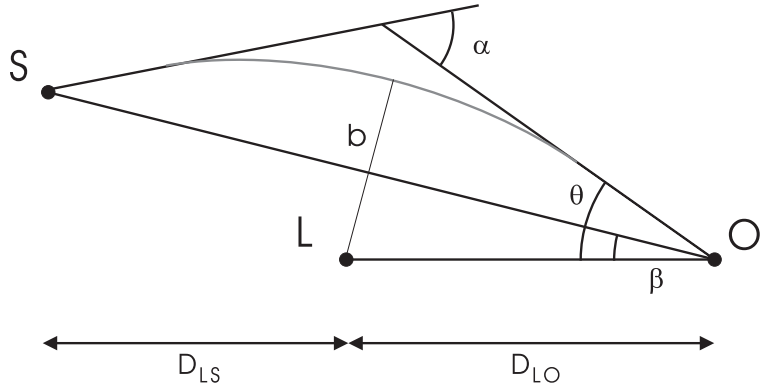}}
\caption{\label{def} The geometry of the lensing event. 
The lens is located at a distance $D_{OL}$ from the observer;
the source, at a distance $D_{SO}$. 
The angular separation between the lens and the source is an angle $ \beta$ and the
position of the image is at an angle $\theta$. The deflection angle is $\alpha$.} 
\end{figure}

This relation is equivalent to:

\begin{eqnarray}
&& \beta = \theta - \frac{\theta_0^2}{\theta} \\
{\rm or} && \beta D_{LO} = b - \frac{b_0^2}{b}
\end{eqnarray}

where $\theta_0, b_0$ are the values of $\theta, b$ in the case of perfect alignment.

The deflection, the characteristic length and angle obey:

\begin{eqnarray}
\begin{array}{rcccl}
\alpha     &=& \frac{4M}{b} g(b) && \label{defl} \\
b_0        &=& \theta_0 D_{LO} &=& \sqrt{ 4M g(b) \cdot \frac{D_{LO} D_{LS}}{D_{SO}} } \label{b0} \\
\theta_0  &=&\frac{b_0}{D_{LO}} &=& \sqrt{ 4M g(b) \cdot \frac{D_{LS}}{D_{LO} D_{SO}} }  
\end{array}
\end{eqnarray}

where $2M = \frac{2GM}{c^2} = R_{Schw.}$ is the Schwartzschild radius of the lens ($M$ is its mass).

If  $R_L < b$ ($R_L$ is the radius of the lens),
we are in the usual (photon) case (OUTcase): the deflection is given by $4M/b$ and  $g(b) =1$.
 
As we have seen in section 2, in general $g(b) \neq 1$ ,and explicit expressions were given for
Newtonian objects, assuming either Gaussian or Lorentzian density profiles.

\bigskip
An interesting quantity is the focal length $f$ of the lens for given $b$:

\begin{eqnarray}
f(b) = \left( \frac{4Mg(b)}{b^2} \right) ^{-1}
\end{eqnarray}

A constant $f(b)$ would signal a perfect lens, where a plane wave focuses in one point.
Solving graphically  $ f= D_{LO}$ determines whether lensing occurs, and what its quality is.

\subsubsection{Alignment}

To observe neutrino lensing will usually require huge amplification, and we will 
see that this only happens in the case of perfect alignment ($\beta =0$). 
An interesting situation occurs when, in a region of $b$, say $[b_{min},b_{max}]$,

\begin{eqnarray}
g(b) = \gamma b^2 \\
\sqrt{4M \gamma \cdot \frac{D_{LO} D_{LS}}{D_{SO}} } \approx 1 
\end{eqnarray}

In this case indeed the deflection angle is proportional to $b$, the focal length 
is nearly constant and the object becomes a good lens: all the neutrinos in that 
region of $b$ are focused on the same point. (The second equation simply 
puts the observer at the focus).

In this situation, all the particles passing through a ring $[b_{min},b_{max}]$ or 
even the full disk if $b_{min} = 0$ focus on the detector. 
The signal enhancement is then given by the ratio of the area of this ring or disk to 
the area of the corresponding disk in case no lensing happens. 
For a Gaussian density profile, the region $[0, r_0]$ has a reasonably good lens behavior.
Slightly outside of the focus area, either a ring or disk would 
in principle be observed, but this requires an angular resolution and statistics unrealistic
for neutrinos; our only hope is to notice the enhancement factor.
\footnote{It is in principle possible to obtain more than one ring,
if $f(b)$ crosses $f=D_{LO}$ more than once, although  
this does not happen for the Gaussian and the Lorentzian density profiles.}

\subsection{Amplification factor.}
\subsubsection{General relations.}

The most general relation for the magnification (here, the signal enhancement)
$\mu$ is \cite{gl}:

\begin{eqnarray}
\label{mutheor}
\mu = {\Delta \Omega \over \Delta \Omega_0} 
= \left| {\rm det} \frac{ \partial {\boldmath \vec \beta}}{\partial {\boldmath \vec \theta}} \right| %%@
^{-1}
= {{\cal A} \over {\cal A}_0}
= \left| \frac{\theta_i \Delta \theta_i }{\beta \Delta \beta} \right|
\end{eqnarray}
 
where $\Delta \Omega$, $\Delta \Omega_0$ stand for the solid angle 
that the particle trajectories span in the sky, respectively with and
without lensing effect;

$\vec \beta$, $\vec \theta$ are the angular coordinates of the source and the image;

$\cal A$, ${\cal A}_0$ are the surfaces of the source and the image referred in a same plane 
(for instance source or lens plane);

The last equality holds because $\vec  \beta$ and $\vec \theta$ are aligned. 
The index $i$ refers to a specific image
(in general there is more than one).

\bigskip
The calculation of the magnification depends on the relation between $\beta$ and $\theta$ (given by Eqs.~
(\ref{angles},\ref{defl})).

\bigskip
In the OUTcase, we have two images and the resulting magnification is well known:

\begin{eqnarray}
\label{mag}
\mu _\pm &=& \frac{1}{4} \left( \frac{\beta}{\sqrt{\beta^2 + 4 \theta _0^2}} +
 \frac{\sqrt{\beta^2 + 4 \theta _0^2}}{\beta} \pm 2 \right) \\
\mu &=& \mu _+ + \mu _- = \frac{\beta^2+2\theta_0^2}{\beta \sqrt{\beta^2 + 4\theta _0^2}} \\
\mu &\stackrel{\beta \rightarrow 0}{\longrightarrow} &\frac{\theta _0}{\beta} \stackrel{def}
{=} \mu^\beta _{OUT}
\end{eqnarray}

For the INcase, no general relation holds 
but we will consider two simplified cases (linear and quadratic):

\begin{eqnarray}
{\rm If} \, g(b)= \gamma b^2 , \, \alpha = 4M \gamma b & \Rightarrow & \mu_i = 
\left( \frac{\theta (\beta)}{\beta}
\right) ^2  \stackrel{def}{=} \mu^{\beta}_{disk} = const.
\end{eqnarray}

A Gaussian density profile, taken close to the center is similar to the quadratic case. 
Note the surprising result that the
magnification is independent of $\beta$. As $g(b) =  \gamma b^2$ 
only between $b=0$ and $b=r_0$ (for the Gaussian
profile), the relation holds for $\beta \leq r_0 / D_{LO}$. 
The well known OUTcase behavior is recovered for $\beta \gg r_0 / D_{LO}$.

\begin{eqnarray}
{\rm If} \, g(b)= \delta b, \, \alpha = 4M\delta \;\;\;& \Rightarrow & \mu_i = 1+ \frac{D_{SL}}{D_{SO}}
\frac{\alpha}{\beta} = 1 + \frac{\theta_0^2}{\theta \beta} 
\stackrel{\beta \rightarrow 0}{\longrightarrow} \frac{\theta_0}{\beta}
\end{eqnarray}

The linear case occurs near $b = r_0$ with both Gaussian and Lorentzian profiles. 
There is only a significant
magnification near alignment. 
Otherwise, the lens only refracts the signal: it just changes the apparent position of the
source.

\subsubsection{Alignment.}

\paragraph{Extended sources.}

\bigskip
This far, we have only considered point-like sources and detectors, and
as a result, their
magnification appears to diverge when source, lens and detector
are aligned Eq.~(\ref{mag}).
For an extended circular source of uniform brightness and of angular radius
$\theta_S$, the magnification factor (OUTcase) is \cite{gl}:

\begin{eqnarray}
\label{magOUTs}
\mu_{OUT}^S &=& \frac{1}{\pi \theta_S^2} \int_0^{\theta_S} 2 \pi \beta d\beta \mu (\beta) \\
      &=& \frac{\sqrt{\theta_S^2 + 4 \theta_0^2}}{\theta_S} \\
      & \stackrel{\theta_S \rightarrow 0}{\longrightarrow} & \frac{2 \theta_0}{\theta_S}
\end{eqnarray}

For the INcase, one evaluates the integral above in the same way, 
using the relations~(\ref{angles},\ref{defl}). 
For the case $g(b) = \gamma b^2 $, i.e. $\alpha= 4M \gamma b$, the image of the source 
(in the lens plane for instance) is a full disk of angular radius 
$\theta _{disk} = r_0 /
D_{LO}$. The amplification factor is then \footnote{
When evaluating the integral, one must substitute
$\int_0^{\theta_S}$ by $\int_0^{\theta_{disk} / \sqrt{\mu^\beta}}$.
This takes into account the finite size of the
lens: the relation $\alpha = 4M \gamma b$ holds only in the interval $[0; \theta_{disk}]$.}

\begin{eqnarray}
\mu_{disk}^S = {{\cal A} \over {\cal A}_0} = \left( \frac{\theta_{disk}}{\theta_S} \right) ^2
\end{eqnarray}

For the case $\alpha = 4M \delta$, during alignment, the image of the source
is a ring of radius $\theta_0$ and width
$\sim \theta_S$. In the limit $\theta_S \rightarrow 0$, 
the amplification is the same as in the OUTcase:

\begin{eqnarray}
\mu^S = {{\cal A} \over {\cal A}_0} = 1+ \frac{2 \theta_0^2}{\theta \theta_S} 
\stackrel{ \theta_S \rightarrow 0}{\longrightarrow} \frac{2 \theta_0}{\theta_S}
\end{eqnarray}

\paragraph{Extended detector.}

\bigskip
Finally, we discuss the case of point-like source aligned with the lens and the observer.
At first glance, the
magnification, $\mu^\beta$ or $ \mu^S$ diverges.
However, we have not yet taken into account the finite size of
the detector. The geometry of our problem is symmetric: one can exchange $\beta$ for $\beta_\o$ 
(the angle
$\widehat{LSO}$), $D_{LS}$ for $D_{LO}$ and $\theta_S$ for $\theta_\o$, 
the latter being the angular radius of the
detector seen from the source. The amplification then reads:

\begin{eqnarray} 
\mu_{OUT}^\o &=& \frac{2 \theta_0}{\theta_\o} \frac{D_{LO}}{D_{LS}} \\
\mu_{disk}^\o &=& \left(  \frac{\theta_{disk}}{\theta_\o}  \frac{D_{LO}}{D_{LS}} \right) ^2
\end{eqnarray}

The factor $D_{LO} / D_{LS}$ is due to the fact that the angle $\theta_\o$ 
is not viewed from the observer but from the
source. When we evaluate the magnification, i.e. the surface of the image with 
and without lensing effect, we have to
refer to a same plane. 
Doing so leads to the factor $D_{LO} / D_{LS}$.

\paragraph{Resulting amplification.}

Until now we have preferred a description with
angles. Maybe the situation is clearer if we translate the expressions into
distances in a given reference plane

\begin{figure}[ht]
\label{dist}
\epsfxsize=10cm
\centerline{\epsffile{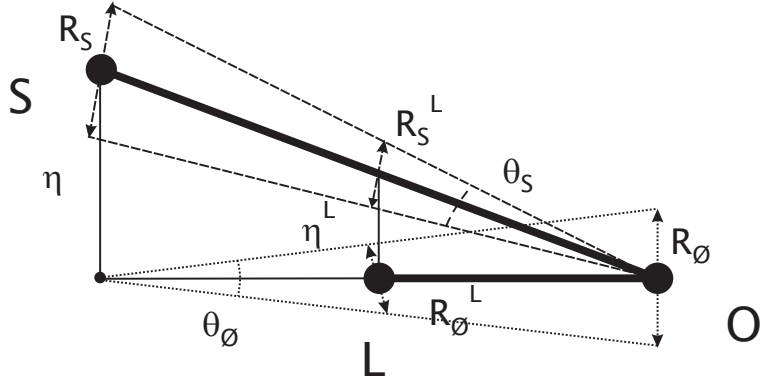}}
\caption{To the angular language corresponds formulations in distances referred to a same plane. 
Here, we refer in the
lens plane. Ex: The source has an angular separation from the lens $\beta$, 
which corresponds to a distance $\eta$ in
the source plane or equivalently $\eta ^L$ in the lens plane.} 
\end{figure}

Choosing to evaluate quantities in the lens plane, $\theta_0 \rightarrow \theta_0 D_{LO} = b_0$, %%@
$\theta_S \rightarrow \theta_S D_{LO} = R_S^L$, the radius of
the source in the lens plane, $\beta \rightarrow \beta D_{LO} = \eta^L$, the position of the source in %%@
the lens
plane, and $\theta_\o D_{LS} /D_{LO} \rightarrow \theta_\o D_{LS} = R_\o^L$, the radius of the detector %%@
in
the lens plane, as seen from the source. The magnification then reads:

\begin{eqnarray}
\begin{array}{c@{\hspace{0.5cm}}c@{\hspace{0.5cm}}c}
\mu_{OUT}^\beta      = \frac{b_0}{\eta^L}  &
\mu_{OUT}^S =\frac{2 b_0}{R_S^L}  &
\mu_{OUT}^\o = \frac{2 b_0}{R_\o^L}  \\
\mu_{disk}^\beta = \left( \frac{b_0}{\eta^L}  \right) ^2 &
\mu_{disk}^S =\left(  \frac{b_{disk}}{R_S^L}  \right) ^2 &
\mu_{disk}^\o = \left(  \frac{b_{disk}}{R_\o^L}   \right) ^2
\end{array}
\end{eqnarray}

As we know by Eq.~(\ref{mutheor}), the magnification factor is given by the ratio $\frac{\cal A}{{\cal %%@
A}_0}$. Which formula to apply  results from a competition between the three angles $\beta$, $\theta_S$
and $\theta_\o D_{LS}/D_{LO}$, or, equivalently, between the lengths $\eta^L$, $R_S^L$ and $R_\o^L$. 
Discarding immediately the off-alignment case, we must compare two situations, using
the lens plane images. If $R_S^L$ (resp. $R_\o^L$) dominates, the image  in the lens plane
is a ring of width $R_S^L$ (resp. $R_\o^L$). Finally, if the image is a
disk, it is independent of the source and the detector {\it during the lensing } but the area of the %%@
image {\it without lensing} is $\pi {R_S^L}^2$ (resp. $\pi {R_\o^L}^2$). So, the answer is: check which %%@
length ($\eta^L$, $R_S^L$ or $R_\o^L$) is largest and use the corresponding expression for the %%@
magnification factor (resp. $\mu^\beta$,  $\mu^S$ or $\mu^\o$). Notice the magnification factor is %%@
inversely proportional to the characteristic length, or its square, so that the smallest value for the %%@
magnification factor is always selected.

Note that in all cases we have the following requirements, in order to use the approximated expressions:

\begin{eqnarray}
\beta , \, \theta_S , \, \theta_\o \, & \ll & \theta_o \\
\eta^L , \, R_S^L , \, R_\o^L \,    & \ll  & b_0
\end{eqnarray}

\subsection{Approximations}

\subsubsection{The distant source:  $D_{SO} \approx D_{SL} \gg D_{LO}$}

If the source is far away, the expression for the impact parameter and the magnification $\mu^\o$ %%@
simplifies:
\mbox{$b_0                \simeq  \sqrt{4M g(b) \, D_{LO} }$},
\mbox{$\theta_0           \simeq  \sqrt{ \frac{4M g(b)}{D_{LO}}}$},
\mbox{$R_\o^L             \simeq  R_\o $}, \linebreak
\mbox{$\mu_{OUT}^\o   \simeq  \frac{2 b_0}{R_\o} $},
\mbox{$\mu_{disk}^\o    \simeq  \left( \frac{b_{disk}}{R_\o}  \right) ^2$}
The following inequalities hold:
\mbox{$\eta^L    \ll  \eta $},
\mbox{$R_S^L  \ll  R_S $}.

A consequence is that in this case one should take care for $R_\o$. Indeed, it is possible that $R_S^L %%@
\ll R_\o$; then
the magnification $\mu^\o$ applies.

\subsubsection{Binary systems: $D_{SO} \approx D_{LO} \gg D_{LS}$}

If the source and the lens are close to each other, as for binary systems \footnote{Henceforth "binary %%@
systems" will
point the case $D_{LO} \gg D_{LS}$} (one object being the lens; the other, the source
\footnote{For a treatment of binary systems, see Ref.~\cite{esteban}}), the impact parameter and the
magnification $\mu^S$ simplifies:
\mbox{$b_0                 \simeq  \sqrt{4M g(b) \, D_{LS} }$},
\mbox{$\theta_0           \simeq  \sqrt{ \frac{4M g(b) \, D_{LS}}{D_{LO}^2}} $},
\mbox{$R_S^L            \simeq R_S$},
\mbox{$\mu_{OUT}^S \simeq \frac{2b_0}{R_S}$} and
\mbox{$\mu_{disk}^S    \simeq  \left( \frac{b_{disk}}{R_S}  \right) ^2 $}.
We have also:
\mbox{$\eta^L    \simeq  \eta $} and
\mbox{$R_\o^L  \ll  R_\o$}.

As $R_\o \leq  1 $ km and as the source is usually macroscopic $R_S \gg 1$ km, it is never needed to take %%@
into account
the finite size of the detector. So $\mu^\o$ never applies. For binary systems the most dangerous %%@
approximation is $R_S
\ll b_0$. This should be checked. If this is not realized, one should use the full expressions, %%@
Eqs.~(\ref{mag},\ref{magOUTs}).

\section{Applications}

We will know concentrate on possible applications and a few examples. In practice, we first test if we %%@
are in an
OUTcase or not. Then we determine the expected shape of the image (discussion of the intersection of %%@
$f(b)$ with
$f=D_{LO}$ and its characteristics. We estimate the magnification by determining the largest length %%@
between
$\eta^L$, $R_S^L$ and $R_\o^L$. We check all these lengths are smaller than $b_0$ to know if we may apply %%@
the
approximations for the magnification.  Finally, we compute the expected signal for a given detector (in %%@
our case
Super-Kamiokande, SK, or IceCube).

\subsection{The Sun}
We consider the Sun as a first example.

\bigskip
Neutrinos passing outside the Sun don't focus close enough to Earth to be of 
use; in fact for them to focus on Earth, we would need

\begin{eqnarray}
b_0 \simeq 30 . 000 {\rm km} \simeq 5 \% R_\odot
\end{eqnarray}
which is clearly well inside the Sun, so the OUT solution does not provide any
sizeable effect.

We turn to the INcase,using as announced a Gaussian density profile as a reasonable
approximation of the matter distribution.
Unfortunately, the plot of the focal length never crosses the line $f=D_{LO}=1$ a.u. The
smallest value is at about 22 a.u., which means Uranians can perform wonderful neutrino lensing %%@
experiments using the
Sun as lens. For them, any source would be amplified in turn as the Sun sweeps in front of it!
(Jupiter cannot replace the Sun as a useful lens for us, as its mass is about $10^{-3} M_\odot$). 

Even if we are not at the focal point, there is yet some amplification due to
the improved convergence of the beam. It provides:

\begin{eqnarray}
\label{nofocus}
A = \left( \frac{f}{ \left| f-D_{LO} \right| } \right) ^2
\end{eqnarray}
The effect is significant if $D_{LO} \simeq f$, say $D_{LO} \in
\left[ 0.3 f ; 3 f \right]$, to have a magnification higher than 2.  
Beyond $3f$, this last effect is negligible.
  
The Sun is thus not a good candidate amplifying a neutrino signal. 
Even if the neutrinos are seriously deflected, the deflection itself is
in  practice impossible to detect, as it has been done for photons, 
this is both due to the insufficient
detection capacity and to the limited angular resolution: $1^\circ$ 
in the best cases, compared to deflections 
of less than one arcminute.

\subsection{Stars}

It is well known that stars are good lens candidates. Indeed, the OUTcase test tells us:

\begin{eqnarray}
b_0 \approx & 20 R_\odot \left( \frac{M}{M_\odot}  \right) ^{1 \over 2}
\left( \frac{D_{LO}}{\rm pc} \right) ^{1 \over 2}
\vspace{4cm}& {\rm if } \; D_{L0} \ll D_{LS} \\
b_0 \approx & 30. 000 {\rm km} \left( \frac{M}{M_\odot}  \right) ^{1 \over 2}
\left( \frac{D_{LS}}{\rm a.u.} \right) ^{1 \over 2}
\vspace{4cm}& {\rm if } \; D_{L0} \gg D_{LS} 
\end{eqnarray}

where $M$ is the mass of the lens and $M_\odot$, $R_\odot$ are the solar mass and radius.

\bigskip
For distant sources, since the closest star is yet at about 1.3 pc, the OUTcase applies for
lensing on stars. The phenomenology is the same as for photons. If the source is aligned, the image is a %%@
ring of radius
$b_0$. The magnification is given by:

\begin{eqnarray}
\begin{array}{ccc@{}c@{}c@{}c@{}c@{\hspace{1.5cm}}c}
\mu^\o_{OUT} & \approx & 13 \times 10^6 & 
\left( \frac{\rm km}{R_\o} \right) & \left( \frac{M}{M_\odot}  \right) ^{1 \over 2} &
\left( \frac{D_{LO}}{\rm pc} \right) ^{1 \over 2} &&  {\rm if} \; R_\o \gg R_s^L \\ 
\mu^S_{OUT} & \approx &39.000  &
\left( \frac{R_\odot}{R_S} \right) & \left( \frac{M}{M_\odot}  \right) ^{1 \over 2} &
\left( \frac{\rm 100 pc}{D_{LO}} \right) ^{1 \over 2 } &\left( \frac{D_{SO}}{\rm 10 kpc} \right) &
{\rm if} \; R_\o \ll R_s^L
\end{array}
\end{eqnarray}

The first line applies when the size of the detector determines the radius of the ring (this never %%@
happens if the
source is a star or a galaxy); the second is typical for a star or a galaxy as
source. However, the "normal" neutrino signal of a star (i.e., assumed to be similar to
the thermonuclear reactions in the Sun) cannot be detected through such 
amplification. In the most favorable
case (short $D_{LO}$ and $D_{SO}$ distances):

\begin{eqnarray*}
& N_{events}=& \left( \frac{L}{L_\odot} \right) \left( \frac{\rm 1 a.u. }{D_{SO}} \right) ^2 \mu_{OUT}^S
\;\; N_\odot \\
% &{\rm For \; a \; star,} \hspace{1.5cm} &\\
&\approx   0.9 \times 10^{-11} & \left( \frac{L}{L_\odot} \right) \left( \frac{R_\odot}{R_S} \right) 
\left( \frac{M}{M_\odot} \right) ^{1 \over 2}  \left( \frac{\rm 100 pc}{D_{SO}} \right)  
\left( \frac{\rm 1 pc}{D_{LO}} \right) ^{1 \over 2} \;\; N_\odot
% &{\rm For \; a \; galaxy,} \hspace{1cm} &\\
%&\approx   0.2 \times 10^{-26}  & \left( \frac{L}{L_\odot} \right) \left( \frac{\rm 100kpc}{R_S} %right) 
%\left( \frac{M}{10^{12} M_\odot} \right) ^{1 \over 2}  \left( \frac{\rm 100 Mpc}{D_{SO}} \right)  
%\left( \frac{D_{LO}}{\rm 1 Mpc} \right) ^{1 \over 2}  \;\; N_\odot
\end{eqnarray*}
The situation could be different if for some reason stellar-size objects happened to be 
strong emitters of energetic neutrinos.
In the case of stellar lenses, galaxies are in general too wide to be used as a neutrino sources: the %%@
magnification is
trifling if the angular radius of the galaxy $\theta_S$ is larger than the angular radius of the ring %%@
$\theta_0$
\footnote{$\theta_S$ is typically larger than $10^{-6}$ for a galaxy and $\theta_0$ is smaller than %%@
$10^{-6}$ for a
star.}.

\subsection{Binary Systems}
We discuss know the INcase, which holds for binary systems. We will concentrate on the most promising %%@
situation, i.e.
 a good lens: as we have seen before, the image of the source is then a full disk. For our
estimations, we have chosen a Gaussian density profile with $r_0 = 0.2 R_L$, the latter giving the %%@
typical size of the
disk. The magnification is thus:

\begin{eqnarray}
\mu^S_{disk} &=& \left( \frac{\theta_{disk}}{\theta_S} \right) ^2 = \left( \frac{r_0}{R_S^L} \right) ^2 %%@
\\
& \approx & \frac{1}{25} \left( \frac{R_L}{R_S} \right) ^2 
\end{eqnarray}

The source should thus be smaller than the lens by at least one order of magnitude.

\bigskip

The very question here is to know if the disk is focused on the observer. If the focal length at $r_0$ is %%@
too short, the disk effect will be negligible, and the effect will
be dominated by an outer part of the lensing star (ring situation); if it is too long, no other part of %%@
the lensing star 
will be able to focus on the signal on the observer,  and the magnification is
given by Eq.~(\ref{nofocus}). Notice here $\alpha = 4M\gamma b^2 \approx 4M ( b/r_0 ) ^2$ and $D_{LO} %%@
\approx D_{SO}$.
We have the focal length:

\begin{eqnarray}
\frac{f_{disk}}{D_{LO}} \approx 22 \left( \frac{R_L}{R_\odot} \right) ^2
\left( \frac{M_\odot}{M} \right)        \left( \frac{\rm 1 a.u.}{D_{LS}} \right)
\end{eqnarray}

Clearly, the result $f_{disk}/ D_{LO} \approx 1$ is achievable. This situation is really promising, %%@
providing the
source is small enough compared to the lens, so that the magnification is significant. Binary stars will %%@
however
seldom meet all the conditions; more exotic systems, with a compact and intense neutrino source, are %%@
needed.

\subsection{Galaxies}
We now turn to another class of lens candidates, namely galaxies or rather 
the galactic halos.
If the source is also galaxy, which is an extended object, we will always be in the INcase. Indeed, from %%@
Eq.~(\ref{b0}),

\begin{eqnarray}
b_0 \leq \sqrt{4M D_{LO}} \approx 14 {\rm kpc} \left( \frac{M}{10^{12} M_\odot} \right) ^{1 \over 2}
\left( \frac{D_{LO}}{\rm Gpc} \right) ^{1 \over 2}
\end{eqnarray}

Now, the radius of galaxies is typically of this length but, since we will focus on the dark matter halo %%@
of the galaxy,
the INcase applies in most situations.

>From Fig.~\ref{plotlorentzian}, the maximum deflection angle stays close to
$4M / R_L$. More precisely, it is always less than $\pi /2 \cdot 4M / R_L$. So,

\begin{eqnarray}
\alpha_{max} \leq 3 \times10^{-6} \left( \frac{M}{10^{12} M_\odot} \right) 
\left( \frac{\rm 100kpc}{R_L} \right) 
\end{eqnarray}

and, in order to focus on Earth,

\begin{eqnarray}
D_{LO} > f_{min} = \frac{b_0}{\alpha_{max}} \approx 30 {\rm Gpc} \left( \frac{b_0}{R_L} \right)
\left( \frac{10^{12} M_\odot}{M} \right) 
\left( \frac{R_L}{\rm 100kpc} \right) 
\end{eqnarray}

Since the radius of the Universe is only about 5 Gpc, the neutrinos must pass rather near the center of %%@
the galaxy,
often through its visible part. They survive however since the probability to meet a star
is tiny. (at the difference of photons, which are absorbed by dust clouds).

For the Lorentzian density profile, the problem is that we have no precise idea of the value of $r_0$, %%@
except that it
is small enough to explain the velocity dispersion curves inside galaxies. Here, we discuss only the case %%@
$b_0 > r_0$.
The deflection angle is then nearly constant: $\alpha \approx 4M /R_L \cdot [ \pi /2 - (\pi/2 - 1) b ]$. %%@
The image is a
ring and the magnification is, providing $R_\o \ll R_s^L$:

\begin{eqnarray}
\nonumber
\mu^S  & \approx & 2.8  
\left( \frac{\rm 100 kpc}{R_S} \right)  \left( \frac{M}{10^{12} M_\odot}  \right) ^{1 \over 2} 
\left( \frac{\rm 10 Mpc}{D_{LO}} \right) ^{1 \over 2 } 
\left( \frac{D_{SO}}{\rm 1 Gpc} \right) ^{1 \over 2 } 
\left( \frac{D_{LS}}{\rm 1 Gpc} \right) ^{1 \over 2 } 
\end{eqnarray}

The magnification and the number of events are
too tiny to be observed. As in the previous section, a small, intense neutrino source is preferable.

\section{Conclusions}
We have studied in some detail the lensing of neutrinos through models 
of astrophysical objects. The main difference with photons rests in
the possibility for medium-energy neutrinos to cross even the core of
stars, and this has the important result that the quality of 
the lens is greatly improved.
This would have been very promising if the focal length of the center of 
the Sun happened to coincide with the radius of the Earth orbit,
unfortunately this is far to be the case, and the focusing occurs closer
to Uranus. 
Provided small and energetic sources exist, binary systems, where one
of the companions is the emitter and the other a larger star, or galaxies
could provide sizable enhancements of  the signal.
 
\section{Acknowledgments}
This work was supported by I.~I.~S.~N. Belgium and by the Communaut\'e 
Fran\c caise de Belgique (Direction de la Recherche Scientifique programme 
ARC).

D.~Monderen and V.~Van Elewyck benefit from a F.~R.~I.~A. grant.

\section*{Appendix}
In this appendix, we present the procedure for calculating the 
deflection of neutrinos from straight-line motion as they pass through a
gravitational field produced by a compact object of mass $M$. Here, we only
consider in detail the case concerning the INside solution, i.e. when 
the neutrino flux passes through the object. 
The procedure is shown for the three
specific density profiles explained in Sec.~\ref{rafel}: constant density,
Gaussian and Lorentzian distribution densities.

The procedure starts with the calculation of the mass and pressure
of the object at a radius $r$, $m(r)$ and $p(r)$ respectively, for a given
density profile $\rho(r)$. We also calculate $\Phi(r)$ and the metric as
explained in the main text. 
Second, we derive from the metric the equation of the
orbit (for the case of constant density we show in detail the derivation) and
the smallest radius reached by the particle. Finally, we calculate the net 
deflection.

We begin by calculating the net deflection for the case of a constant density
profile $\rho(r)=\rho$
\begin{equation}
\label{INconstrho}
\begin{array}{l}
\Longrightarrow\ m(r)=\frac{4\pi}{3}\rho r^3=M\left(\frac{r}{R}\right)^3\\[1ex]
\Longrightarrow\ p(r)=\rho \frac{\sqrt{1-\frac{2M}{R}\frac{r^2}{R^2}}-
                            \sqrt{1-\frac{2M}{R}}}
                           {3\sqrt{1-\frac{2M}{R}}-
                            \sqrt{1-\frac{2M}{R}\frac{r^2}{R^2}}}\\[2ex]
\Longrightarrow\ e^{2\Phi(r)}=
                 \left(\frac{3}{2}\sqrt{1-\frac{2M}{R}}-
                       \frac{1}{2}\sqrt{1-\frac{2M}{R}\frac{r^2}{R^2}}
                 \right)^2\\[1ex]
\Longrightarrow\
ds^2=-\left(\frac{3}{2}\sqrt{1-\frac{2M}{R}}-
            \frac{1}{2}\sqrt{1-\frac{2M}{R}\frac{r^2}{R^2}}\right)^2 dt^2+
\frac{1}{1-\frac{2M}{r}\left(\frac{r}{R}\right)^3}dr^2+r^2 d\Omega^2\ . 
\end{array}
\end{equation}
For a massless particle
\begin{equation}
\label{eqdiforbitINconstrhoprep}
\left.
\begin{array}{l}
p_0=-E\\[1ex]
p^r=\frac{dr}{d\lambda}\\[2ex]
p_\phi=L
\end{array}
\right\}\ \Longrightarrow\ 
\left\{
\begin{array}{l}
p^0=\frac{E}{\left(\frac{3}{2}\sqrt{1-\frac{2M}{R}}-
                   \frac{1}{2}\sqrt{1-\frac{2M}{R}\frac{r^2}{R^2}}\right)^2}
\\[2ex]
p_r=\frac{1}{1-\frac{2M}{r}\left(\frac{r}{R}\right)^3}
    \frac{dr}{d\lambda}\\[2ex]
p^\phi=\frac{d\phi}{d\lambda}=\frac{L}{r^2}
\end{array}
\right.
\end{equation}
and the equation of the orbit ($p^2=0$) is then
\begin{equation}
\label{eqdiforbitINconstrho}
\begin{array}{rl}
\frac{d\phi}{dr}=&
\frac{\frac{3}{2}\sqrt{1-\frac{2M}{R}}-
      \frac{1}{2}\sqrt{1-\frac{2M}{R}\frac{r^2}{R^2}}}
     {\sqrt{1-\frac{2M}{r}\left(\frac{r}{R}\right)^3}}
\frac{1}{r^2\sqrt{\frac{1}{b^2}-\frac{1}{r^2}
\left(\frac{3}{2}\sqrt{1-\frac{2M}{R}}-
      \frac{1}{2}\sqrt{1-\frac{2M}{R}\frac{r^2}{R^2}}\right)^2}}\\[2ex]
\Longrightarrow &
\frac{d\phi}{du}=
\frac{\left(1-\frac{3M}{2R}\right)+\frac{3M}{2R^3}\frac{1}{u^2}}
     {\sqrt{\left(\frac{1}{b^2}-\frac{M}{R^3}\right)-u^2
            \left(1-\frac{3M}{R}\right)}}+O(M^2 u^2)\ .
\end{array}
\end{equation}
Integrating Eq.~(\ref{eqdiforbitINconstrho}) gives
\begin{equation}
\label{phiINuconstrho}
\begin{array}{rl}
\phi_{\rm IN}(u)=&\phi_0+\frac{2M}{b}
\left(1-\sqrt{1-\frac{b^2}{R^2}}\right)
+\arcsin\left[\frac{b}{R}\left(1-\frac{M}{R}\right)\right]\\[2ex]
+&\frac{3M}{2R}\frac{b^2}{R^2}\left(\sqrt{1-\frac{b^2}{R^2}}\frac{R}{b}-
                                    \sqrt{1-b^2 u^2}\frac{1}{b
u}\right)\\[2ex]
-&\arcsin\left\{\frac{b}{R}\left[1-\frac{3M}{2R}
         \left(1-\frac{b^2}{3R^2}\right)\right]\right\}\\[2ex]
+&\arcsin\left\{b u\left[1-\frac{3M}{2R}
         \left(1-\frac{b^2}{3R^2}\right)\right]\right\}\ ,
\end{array}
\end{equation}
where the integration constant is defined so as 
$\phi_{\rm IN}(r=R)=\phi_{\rm OUT}(r=R)$ (with $\phi_0$ the initial 
incoming direction).
The particle reaches its smallest $r$ when $\frac{dr}{d\lambda}=0$:
\begin{equation}
\label{eqdifrminINconstrho}
\begin{array}{rl}
\frac{dr}{d\lambda}=&E\sqrt{\left[
\frac{1}{\left(\frac{3}{2}\sqrt{1-\frac{2M}{R}}-
               \frac{1}{2}\sqrt{1-\frac{2M}{R}\frac{r^2}{R^2}}\right)^2}-
\frac{b^2}{r^2}\right]
\left[1-\frac{2M}{r}\left(\frac{r}{R}\right)^3\right]}
=0\\[2ex]
\Longrightarrow&
u_{\rm max}=\frac{1}{b}\left[1+\frac{3M}{2R}\left(1-\frac{b^2}{3R^2}\right)
\right]+O(M^2 u^2)\ ,
\end{array}
\end{equation}
then
\begin{equation}
\label{phiINumax}
\begin{array}{rl}
\phi_{\rm IN}(u=u_{\rm max})=&
\phi_0+\frac{\pi}{2}+\frac{2M}{b}
\left(1-\sqrt{1-\frac{b^2}{R^2}}\right)
+\arcsin\left[\frac{b}{R}\left(1-\frac{M}{R}\right)\right]\\[2ex]
+&\frac{3M}{2R}\frac{b}{R}\sqrt{1-\frac{b^2}{R^2}}
-\arcsin\left\{\frac{b}{R}\left[1-\frac{3M}{2R}
        \left(1-\frac{b^2}{3R^2}\right)\right]\right\}\ ,
\end{array}
\end{equation}
and the net deflection is
\begin{equation}
\label{phiINnetconstrho}
\begin{array}{rl}
\Delta\phi=&
\left\{
\begin{array}{ll}
\frac{4M}{b} & {\rm if}\ b\geq R\\[2ex]
\frac{4M}{b}\left(1-\sqrt{1-\frac{b^2}{R^2}}\right)
+2\arcsin\left[\frac{b}{R}\left(1-\frac{M}{R}\right)\right]\\[2ex]
\ \ +\frac{3M}{R}\frac{b}{R}\sqrt{1-\frac{b^2}{R^2}}
    -2\arcsin\left\{\frac{b}{R}\left[1-\frac{3M}{2R}
     \left(1-\frac{b^2}{3R^2}\right)\right]\right\} & {\rm if}\ b<R
\end{array}
\right. \\[8ex]
=&\left\{
\begin{array}{ll}
\frac{4M}{b} & {\rm if}\ b>R\\[2ex]
\frac{4M}{R} & {\rm if}\ b=R\\[2ex]
\frac{4M}{R}\frac{3b}{2R} & {\rm if}\ b\ll R\\[2ex]
0 & {\rm if}\ b=0
\end{array}
\right.
\end{array}
\end{equation}
where the outside solution is also included for completeness.

Next, we present the analysis for the Gaussian and Lorentzian
distribution densities. As explained in the main text, in both cases, it is 
possible to make some approximations 
(see Eqs.~(\ref{PhirlimNewton},\ref{metriclimNewton}) in Sec.~\ref{rafel}) 
that allow us to derive an analytical expression for the net deflection. 
For a Gaussian distribution density profile $\rho(r)=\rho_0 e^{-r^2/r_0^2}$
\begin{equation}
\begin{array}{rl}
\Longrightarrow &
m(r)=M\frac{r}{R}e^{(R^2-r^2)/r_0^2}
\frac{r_0/r\,e^{r^2/r_0^2}\sqrt\pi/2\,
      {\rm erf}\left(r/r_0\right)-1}
     {r_0/R\,e^{R^2/r_0^2}\sqrt\pi/2\,
      {\rm erf}\left(R/r_0\right)-1}\\[2ex]
\Longrightarrow &
\Phi(r)=-\frac{M}{R}
\frac{r_0/r\,e^{R^2/r_0^2}\sqrt\pi/2\,
      {\rm erf}\left(r/r_0\right)-1}
     {r_0/R\,e^{R^2/r_0^2}\sqrt\pi/2\,
      {\rm erf}\left(R/r_0\right)-1}\\[2ex]
\Longrightarrow & 
ds^2=-\left(1-\frac{2M}{R}
\frac{r_0/r\,e^{R^2/r_0^2}\sqrt\pi/2\,
      {\rm erf}\left(r/r_0\right)-1}
     {r_0/R\,e^{R^2/r_0^2}\sqrt\pi/2\,
      {\rm erf}\left(R/r_0\right)-1}\right)dt^2\\[2ex]
& +\left(1+\frac{2M}{R}e^{(R^2-r^2)/r_0^2}
\frac{r_0/r\,e^{r^2/r_0^2}\sqrt\pi/2\,
      {\rm erf}\left(r/r_0\right)-1}
     {r_0/R\,e^{R^2/r_0^2}\sqrt\pi/2\,
      {\rm erf}\left(R/r_0\right)-1}\right)dr^2+r^2 d\Omega^2\ ,
\end{array}
\end{equation}
where the error function is defined as 
${\rm erf}(z)=\frac{2}{\sqrt\pi}\int_0^z dt\,e^{-t^2}$.
The equation of the orbit is
\begin{equation}
\label{eqdiforbitIN}
\begin{array}{l}
\frac{d\phi}{dr}=
\left[1-\frac{M}{R}
\frac{e^{(R^2-r^2)/r_0^2}-1}
     {r_0/R\,e^{R^2/r_0^2}\sqrt\pi/2\,
      {\rm erf}\left(R/r_0\right)-1}\right]
\frac{1}{r^2\sqrt{\frac{1}{b^2}-\frac{1}{r^2}
\left[1-\frac{2M}{R}
\frac{r_0/r\,e^{R^2/r_0^2}\sqrt\pi/2\,
      {\rm erf}\left(r/r_0\right)-1}
     {r_0/R\,e^{R^2/r_0^2}\sqrt\pi/2\,
      {\rm erf}\left(R/r_0\right)-1}\right]}}\\[4ex]
\Longrightarrow\ 
\frac{d\phi}{dy}=
\left\{1+\frac{2M}{R}
\frac{e^{R^2/r_0^2}\left[
\sqrt\pi/2\,y r_0\,{\rm erf}\left(1/y r_0\right)-e^{-1/(y
r_0)^2}\right]}
{r_0/R\,e^{R^2/r_0^2}\sqrt\pi/2\,{\rm erf}\left(R/r_0\right)-1}
\right\}
\frac{1}{\sqrt{\frac{1}{b^2}-y^2}}+O(M^2 u^2)\ ,
\end{array}
\end{equation}
where $y$ is defined as
\begin{equation}
y\equiv \frac{1}{r}\left[1-\frac{M}{R}
\frac{r_0/r e^{R^2/r_0^2}\sqrt\pi/2\,
      {\rm erf}\left(r/r_0\right)-1}
     {r_0/R\,e^{R^2/r_0^2}\sqrt\pi/2\,
      {\rm erf}\left(R/r_0\right)-1}\right]\ .
\end{equation}
Integrating Eq.~(\ref{eqdiforbitIN}) gives
\begin{equation}
\label{phiIN}
\begin{array}{rl}
\phi_{\rm IN}(y)=&
\phi_0+\frac{2M}{b}\left(1-\sqrt{1-\frac{b^2}{R^2}}\right)+\arcsin(by)\\[2ex]
+&\frac{2M}{b}\frac{r_0/R\,e^{R^2/r_0^2}\sqrt\pi/2}
{r_0/R\,e^{R^2/r_0^2}\sqrt\pi/2\,{\rm erf}\left(R/r_0\right)-1}\\[2ex]
\times &\left\{\sqrt{1-\frac{b^2}{R^2}}\,{\rm erf}\left(R/r_0\right)-
         \sqrt{1-b^2 y^2}\,{\rm erf}\left(1/y r_0\right)\right.\\[2ex]
&\left. -e^{-b^2/r_0^2}
\left[{\rm erf}\left(\sqrt{1-\frac{b^2}{R^2}}\frac{R}{r_0}\right)-
      {\rm erf}\left(\frac{\sqrt{1-b^2 y^2}}{y r_0}\right)\right]\right\}\ .
\end{array}
\end{equation}
The particle reaches its smallest $r$ at $y_{\rm max}=\frac{1}{b}+O(M^2 u^2)$,
then
\begin{equation}
\label{phiINymax}
\begin{array}{rl}
\phi(y=\frac{1}{b})=&
\phi_0+\frac{\pi}{2}+\frac{2M}{b}\left(1-\sqrt{1-\frac{b^2}{R^2}}\right)
+\frac{2M}{b}\frac{r_0/R\,e^{R^2/r_0^2}\sqrt\pi/2}
{r_0/R\,e^{R^2/r_0^2}\sqrt\pi/2\,{\rm erf}\left(R/r_0\right)-1}\\[2ex]
\times &\left[ \sqrt{1-\frac{b^2}{R^2}}\,{\rm erf}\left(R/r_0\right)-
e^{-b^2/r_0^2}{\rm erf}\left(\sqrt{1-\frac{b^2}{R^2}}\frac{R}{r_0}\right)
\right] \ ,
\end{array}
\end{equation}
and the net deflection is
\begin{equation}
\label{phiINnetgaussian}
\begin{array}{rl}
\Delta\phi=&
\left\{
\begin{array}{ll}
\frac{4M}{b} & {\rm if}\ b\geq R\\[2ex]
\frac{4M}{b}\left(1-\sqrt{1-\frac{b^2}{R^2}}\right)
+\frac{4M}{b}\frac{r_0/R\,e^{R^2/r_0^2}\sqrt\pi/2}
{r_0/R\,e^{R^2/r_0^2}\sqrt\pi/2\,{\rm erf}\left(R/r_0\right)-1}\\[2ex]
\times\left\{\sqrt{1-\frac{b^2}{R^2}}\,{\rm erf}\left(R/r_0\right)-
e^{-b^2/r_0^2}{\rm erf}\left(\sqrt{1-\frac{b^2}{R^2}}\frac{R}{r_0}\right)
\right\} & {\rm if}\ b<R
\end{array}
\right. \\[8ex]
=&\left\{
\begin{array}{ll}
\frac{4M}{b} & {\rm if}\ b>R\\[2ex]
\frac{4M}{R} & {\rm if}\ b=R\\[2ex]
\frac{4M}{R}\frac{e^{R^2/r_0^2}\sqrt\pi/2\,{\rm erf}\left(R/r_0\right)}
{r_0/R\,e^{R^2/r_0^2}\sqrt\pi/2\,{\rm erf}\left(R/r_0\right)-1}
\frac{b}{r_0} & {\rm if}\ b\ll R\\[2ex]
0 & {\rm if}\ b=0
\end{array}
\right.
\end{array}
\end{equation}
For a Lorentzian distribution density profile 
$\rho(r)=\frac{\rho_0}{1+r^2/r_0^2}$
\begin{equation}
\begin{array}{rl}
\Longrightarrow &
m(r)=M\frac{r}{R}
\frac{1-r_0/r\,\arctan(r/r_0)}
     {1-r_0/R\,\arctan(R/r_0)}\\[2ex]
\Longrightarrow &
\Phi(r)=-\frac{M}{R}
\frac{1-r_0/r\,\arctan(r/r_0)-\frac{1}{2} 
      \log\left(\frac{r^2+r_0^2}{R^2+r_0^2}\right)}
     {1-r_0/R\,\arctan(R/r_0)}\\[2ex]
\Longrightarrow & 
ds^2=-\left(1-\frac{2M}{R}
\frac{1-r_0/r\,\arctan(r/r_0)-\frac{1}{2}
      \log\left(\frac{r^2+r_0^2}{R^2+r_0^2}\right)}
     {1-r_0/R\,\arctan(R/r_0)}\right)dt^2\\[2ex]
& +\left(1+\frac{2M}{R}
\frac{1-r_0/r\,\arctan(r/r_0)}
     {1-r_0/R\,\arctan(R/r_0)}\right)dr^2+r^2 d\Omega^2\ .
\end{array}
\end{equation}
The equation of the orbit is
\begin{equation}
\label{eqdiforbitINlorentzian}
\begin{array}{l}
\frac{d\phi}{dr}=
\left[1+\frac{M}{R}\frac{\frac{1}{2}\log\left(\frac{r^2+r_0^2}{R^2+r_0^2}\right)}
                        {1-r_0/R\,\arctan(R/r_0)}\right]
\frac{1}{r^2\sqrt{\frac{1}{b^2}-\frac{1}{r^2}
\left[1-\frac{2M}{R}
\frac{1-r_0/r\,\arctan(r/r_0)-\frac{1}{2} 
      \log\left(\frac{r^2+r_0^2}{R^2+r_0^2}\right)}
     {1-r_0/R\,\arctan(R/r_0)}\right]}}\\[4ex]
\Longrightarrow\ 
\frac{d\phi}{dy}=
\left[1+\frac{2M}{R}
\frac{1-r_0 y\,\arctan(1/r_0 y)}{1-r_0/R\,\arctan(R/r_0)}\right]
\frac{1}{\sqrt{\frac{1}{b^2}-y^2}}+O(M^2 u^2)\ ,
\end{array}
\end{equation}
where $y$ is defined as
\begin{equation}
y\equiv \frac{1}{r}\left[1-\frac{M}{R}
\frac{1-r_0/r\,\arctan(r/r_0)- \frac{1}{2}
      \log\left(\frac{r^2+r_0^2}{R^2+r_0^2}\right)}
     {1-r_0/R\,\arctan(R/r_0)}\right]\ .
\end{equation}
Integrating Eq.~(\ref{eqdiforbitINlorentzian}) gives
\begin{equation}
\label{phiINlorentzian}
\begin{array}{rl}
\phi_{\rm IN}(y)=&
\phi_0+\frac{2M}{b}\left(1-\sqrt{1-\frac{b^2}{R^2}}\right)+\arcsin(by)
-\frac{2M}{b}\frac{1}{1-r_0/R\,\arctan(R/r_0)}\frac{r_0}{R}\\[2ex]
\times &\left\{\sqrt{1-\frac{b^2}{R^2}}\arctan(R/r_0)-
               \sqrt{1-b^2 y^2}\arctan(1/r_0 y)\right.\\[2ex]
&\left. -\sqrt{1+\frac{b^2}{r0^2}}
\left[\arctan\left(\frac{\sqrt{1-\frac{b^2}{R^2}}}{\sqrt{1+\frac{b^2}{r0^2}}}
                   \frac{R}{r_0}\right)-
      \arctan\left(\frac{\sqrt{1-b^2 y^2}}{\sqrt{1+\frac{b^2}{r0^2}}}
                   \frac{1}{r_0 y}\right)\right]\right\}\ .
\end{array}
\end{equation}
The particle reaches its smallest $r$ at $y_{\rm max}=\frac{1}{b}+O(M^2 u^2)$,
then
\begin{equation}
\label{phiINymaxlorentzian}
\begin{array}{rl}
\phi(y=\frac{1}{b})=&
\phi_0+\frac{\pi}{2}+\frac{2M}{b}\left(1-\sqrt{1-\frac{b^2}{R^2}}\right)
-\frac{2M}{b}\frac{1}{1-r_0/R\,\arctan(R/r_0)}\frac{r_0}{R}\\[2ex]
\times &\left[\sqrt{1-\frac{b^2}{R^2}}\arctan(R/r_0)
              -\sqrt{1+\frac{b^2}{r0^2}}
\arctan\left(\frac{\sqrt{1-\frac{b^2}{R^2}}}{\sqrt{1+\frac{b^2}{r0^2}}}
                   \frac{R}{r_0}\right)\right]\ ,
\end{array}
\end{equation}
and the net deflection is
\begin{equation}
\label{phiINnetlorentzian}
\begin{array}{rl}
\Delta\phi=&
\left\{
\begin{array}{ll}
\frac{4M}{b} & {\rm if}\ b\geq R\\[2ex]
\frac{4M}{b}\left(1-\sqrt{1-\frac{b^2}{R^2}}\right)
-\frac{4M}{b}\frac{1}{1-r_0/R\,\arctan(R/r_0)}\frac{r_0}{R}\\[2ex]
\times\left[
\begin{array}{l}
\sqrt{1-\frac{b^2}{R^2}}\arctan(R/r_0)\\[2ex]
-\sqrt{1+\frac{b^2}{r0^2}}
\arctan\left(\frac{\sqrt{1-\frac{b^2}{R^2}}}{\sqrt{1+\frac{b^2}{r0^2}}}
             \frac{R}{r_0}\right)
\end{array}
\right] & {\rm if}\ b<R
\end{array}
\right. \\[12ex]
=&\left\{
\begin{array}{ll}
\frac{4M}{b} & {\rm if}\ b>R\\[2ex]
\frac{4M}{R} & {\rm if}\ b=R\\[2ex]
\frac{2M}{R}\frac{\arctan(R/r_0)}{1-r_0/R\,\arctan(R/r_0)}
\frac{b}{r_0} & {\rm if}\ b\ll R\\[2ex]
0 & {\rm if}\ b=0
\end{array}
\right.
\end{array}
\end{equation}

\end{document}